\documentclass{article}
\usepackage{epsfig}
\usepackage{amssymb}

\newcommand{\be}{\begin{equation}}
\newcommand{\ee}{\end{equation}}
\newcommand{\ba}{\begin{eqnarray}}
\newcommand{\ea}{\end{eqnarray}}

\begin{document}

\title{Cooper pairing reexamined}
\author{M. Fortes,$^{a}$ M. de Llano,$^{b}$ and M. A. Sol\'{\i}s,$^{a}$ \\
$^{a}$Instituto de F\'{\i}sica, Universidad Nacional Aut\'{o}noma de M\'{e}%
xico, \\ Apdo. Postal 20-364, 01000 M\'{e}xico, DF, Mexico \\
$^{b}$Instituto de Investigaciones en Materiales, Universidad Nacional Aut%
\'{o}noma de M\'{e}xico, \\Apdo. Postal 70-360, 04510 M\'{e}xico, DF, Mexico\\
}
\maketitle

\begin{abstract}
When both two-electron \textit{and} two-hole Cooper-pairing are treated on
an equal footing in the ladder approximation to the Bethe-Salpeter (BS)
equation, the zero-total-momentum Cooper-pair energy is found to have two 
\textit{real} solutions $\mathcal{E}_{0}^{BS}=\pm 2\hbar \omega _{{D}%
}/\sqrt{{e}^{2/\lambda }+{1}}$ which coincide with the
zero-temperature BCS energy gap $\Delta =\hbar \omega _{D}/\sinh (1/\lambda )
$ in the weak coupling limit. Here, $\hbar \omega _{D}$ is the Debye energy
and $\lambda \geq 0$ the BCS model interaction coupling parameter. The
interpretation of the BCS energy gap as the binding energy of a Cooper-pair
is often claimed in the literature but, to our knowledge, never
substantiated even in weak-coupling as we find here. In addition, we confirm
the two purely-\textit{imaginary} solutions assumed since at least the late
1950s as the \textit{only} solutions, namely, $\mathcal{E}_{0}^{BS}=\pm
i2\hbar \omega _{{D}}/\sqrt{{e}^{2/\lambda }{-1}}.$
\end{abstract}

The bound-state, two-particle Bethe-Salpeter (BS) \cite{BS}\ wavefunction in
the ladder approximation, with both particle- and hole-propagation, for the
ideal Fermi gas (IFG)-based generalized Cooper pair (CP) problem\ \cite%
{Honolulu}\ is%
\begin{eqnarray}
\psi (\mathbf{k,}E) &=&-\left( \frac{i}{\hbar }\right) ^{2}G_{0}\left( 
\mathbf{K}/2+\mathbf{k},\mathcal{E}_{K}/2+E\right) G_{0}\left( \mathbf{K}/2-%
\mathbf{k},\mathcal{E}_{K}/2-E\right) \times  \nonumber \\
&&\frac{1}{2\pi i}\int\limits_{-\infty }^{+\infty }dE^{^{\prime }}\frac{1}{%
L^{d}}\sum_{\mathbf{k}^{\prime }}v(\left\vert \mathbf{k-k}^{\prime
}\right\vert )\psi (\mathbf{k}^{\prime },E^{\prime }).  \label{BSE}
\end{eqnarray}%
Here $L^{d}$ is the \textquotedblleft volume\textquotedblright\ of the $d$%
-dimensional system; $\mathbf{K\equiv k}_{1}+\mathbf{k}_{2}$ is the total or
center-of-mass momentum (CMM) and $\mathbf{k\equiv 
{\frac12}%
(k}_{1}-\mathbf{k}_{2}\mathbf{)}$ the relative momentum wavevectors of the
two-particle bound state whose wavefunction is $\psi (\mathbf{k,}E)$; $%
v(\left\vert \mathbf{k-k}^{\prime }\right\vert )$ is the Fourier transform
of the interparticle interaction, $\mathcal{E}_{K}$ $\equiv E_{1}+E_{2}$ is
the energy of this bound state while $E\equiv E_{1}-E_{2}$, and $G_{0}\left( 
\mathbf{K}/2+\mathbf{k},\mathcal{E}/2+E\right) $ is the bare one-fermion
Green's function given by (\cite{FW}, p. 72)%
\begin{equation}
G_{0}(\mathbf{k}_{1},E_{1})=\frac{\hbar }{i}\left\{ \frac{\theta
(k_{1}-k_{F})}{-E_{1}+\epsilon _{\mathbf{k}_{1}}-E_{F}-i\varepsilon }+\frac{%
\theta (k_{F}-k_{1})}{-E_{1}+\epsilon _{\mathbf{k}_{1}}-E_{F}+i\varepsilon }%
\right\}  \label{IFGgreen}
\end{equation}%
where $\epsilon _{\mathbf{k}_{1}}\equiv $ $\hbar ^{2}k_{1}^{2}/2m$ and $%
\theta (x)$ is the step function, so that the first term refers, e.g., to 
\textit{electrons} and the second to \textit{holes}. The latter are also
fermions but of positive charge $+e$.

Consider first the case where holes are ignored, i.e., neglect the second
term in (\ref{IFGgreen}). Note that the energy dependence in (\ref{BSE})
derives from the Green's function only and therefore allows defining a new
function $\varphi _{\mathbf{k}}$ by first writing 
$$\psi (\mathbf{k,}E)\equiv
G_{0}\left( \mathbf{K}/2+\mathbf{k},\mathcal{E}_{K}/2+E\right) G_{0}\left( 
\mathbf{K}/2-\mathbf{k},\mathcal{E}_{K}/2-E\right) \varphi _{\mathbf{k}}$$
which upon substitution in (\ref{BSE}) yields%
\begin{eqnarray}
\varphi _{\mathbf{k}} &=& -\frac{1}{L^{d}}\sum_{\mathbf{k}^{\prime
}}v(\left\vert \mathbf{k-k}^{\prime }\right\vert )\varphi (\mathbf{k}%
^{\prime })\left( \frac{i}{\hbar }\right) ^{2}\frac{1}{2\pi i}\times  \nonumber
\\
&& \int\limits_{-\infty }^{+\infty }dE^{\prime }\frac{\theta (k^{\prime
}-k_{F})}{-\mathcal{E}_{K}/2-E^{\prime }+\epsilon _{\mathbf{K/}2+\mathbf{k}%
^{\prime }}-E_{F}-i\varepsilon }\frac{\theta (k^{\prime }-k_{F})}{-\mathcal{E%
}_{K}/2+E^{\prime }+\epsilon _{\mathbf{K/}2-\mathbf{k}^{\prime
}}-E_{F}-i\varepsilon }.  \label{phi}
\end{eqnarray}%
The energy-$E^{\prime }$ integration then leaves%
\begin{equation}
\varphi _{\mathbf{k}}=-\frac{1}{L^{d}}\sum_{\mathbf{k}^{\prime
}}v(\left\vert \mathbf{k-k}^{\prime }\right\vert )\frac{\theta (k^{\prime
}-k_{F})}{\epsilon _{\mathbf{K/}2+\mathbf{k}^{\prime }}+\epsilon _{\mathbf{K/%
}2-\mathbf{k}^{\prime }}+2E_{F}-\mathcal{E}_{K}}\varphi _{\mathbf{k}^{\prime
}}  \label{BG}
\end{equation}%
which may be recognized as the Bethe-Goldstone (BG) equation \cite{BG}; see
Fig. 1 below.

\begin{figure}[tbh]
\centerline{\epsfig{file=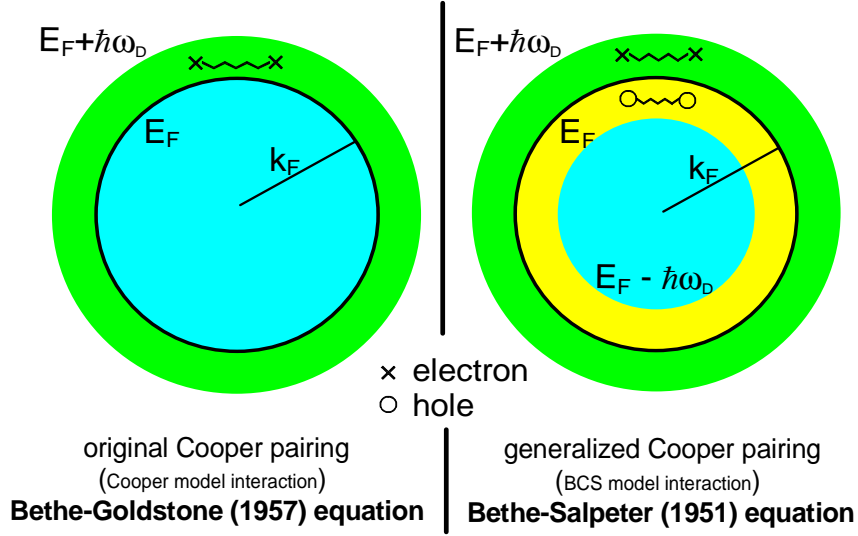,height=3.in,width=5.0in}}
\caption{ The Bethe-Goldstone equation (left) considers only 2e-CPs.
The more general BS (right) includes 2h-CPs.}
\end{figure}

For the ideal-Fermi-gas-sea-based scenario for CPs when holes are \textit{not%
} neglected, we assume the BCS model interaction $v(\left\vert \mathbf{k-k}%
^{\prime }\right\vert )=-V\theta _{BCS}(\epsilon _{\mathbf{k}})\theta
_{BCS}(\epsilon _{\mathbf{k}^{\prime }})$\ where $V\geq 0$ is the strength
of the net attraction between pair partners and the unit step functions $%
\theta _{BCS}(\epsilon )$ restrict particle or hole energies $\epsilon
_{k_{1}},$ $\epsilon _{k_{2}}$ to an energy interval of width $2\hbar \omega
_{D}$ around the Fermi level $E_{F}$, namely $E_{F}\leq \epsilon _{\mathbf{k}%
_{1}},\epsilon _{\mathbf{k}_{2}}\leq E_{F}+\hbar \omega _{D}$ (for
particles) and $E_{F}-\hbar \omega _{D}\leq \epsilon _{\mathbf{k}%
_{1}},\epsilon _{\mathbf{k}_{2}}\leq E_{F}$ (for holes). Integration over
energies in (\ref{phi}) can then be evaluated directly in the complex $%
E^{\prime }$-plane resulting in the following equation for the wavefunction $%
\varphi _{\mathbf{k}}$ with \textit{zero}\ CMM $\mathbf{K}=0$ that is%
\begin{equation}
\varphi _{\mathbf{k}}=\frac{V}{L^{d}}\sum_{\mathbf{k}^{\prime }}{}^{^{\prime
}}\frac{1}{(2\epsilon _{\mathbf{k}^{\prime }}-2E_{F}-\mathcal{E}_{0})}%
\varphi _{\mathbf{k}^{\prime }}-\frac{V}{L^{d}}\sum_{\mathbf{k}^{\prime
}}{}^{^{\prime \prime }}\frac{1}{(2\epsilon _{\mathbf{k}^{\prime }}-2E_{F}-%
\mathcal{E}_{0})}\varphi _{\mathbf{k}^{\prime }}  \label{CP}
\end{equation}%
where $\mathcal{E}_{0}$ is the $\mathbf{K}=0$\ eigenvalue energy. The single
prime over the first (2e-CP) summation term denotes the restriction $%
E_{F}<\epsilon _{\mathbf{k}^{\prime }}<E_{F}+\hbar \omega _{D}$ while the
double prime in the last (2h-CP) term means $E_{F}-\hbar \omega
_{D}<\epsilon _{\mathbf{k}^{\prime }}<E_{F}$. The ordinary Cooper (or BG)
problem is compared in Fig. 1 with the BS problem where electron-hole
symmetry is restored through inclusion of 2h-CPs, represented by the second
term of (\ref{CP}), in addition to the 2e-CPs. Ignoring the second term of (%
\ref{CP}) gives the well-known solution \cite{Coo}%
\begin{equation}
\mathcal{E}_{0}^{C}=-2\hbar \omega_{D}/(e^{2/\lambda }-1) \quad %
\mathrel{\mathop{\longrightarrow}\limits_{\lambda \rightarrow 0}} \quad
-2\hbar \omega _{D}\exp (-2/\lambda )  \label{ordCP}
\end{equation}%
corresponding to a negative-energy, stationary-state bound 2e-CP, where $%
\lambda \equiv VN(E_{F})\geq 0$ with $N(E_{F})$ the electronic density of
states for one spin. Note that a 2e-CP state for general CMM wavevector $%
\mathbf{K}$ of energy $\mathcal{E}_{K}^{C}$\ is characterized only by a
definite $K$ but \textit{not }definite relative-momentum wavevector. This
alone implies that ordinary, as well as the generalized CPs to be considered
below, obey Bose statistics \cite{BCSandCPs,PRBcomm}. Without the first
summation term in (\ref{CP})\ the same expression for the $\mathcal{E}%
_{0}^{C}$\ of 2e-CPs follows for 2h-CPs, apart from an overall sign change.

Since (\ref{CP}) includes 2h-CPs along with 2e-CPs, eliminating the $\varphi
_{\mathbf{k}}$'s from (\ref{CP}) leads to the BS eigenvalue equation for the
pair energy $\mathcal{E}_{0}$ 
\begin{equation}
2/\lambda \, \, =\int\limits_{E_{F}}^{E_{F}+\hbar \omega _{D}}\frac{%
d\epsilon }{\epsilon -E_{F}-\mathcal{E}_{0}/2} \, \,
-\int\limits_{E_{F}-\hbar \omega _{D}}^{E_{F}}\frac{d\epsilon }{\epsilon
-E_{F}-\mathcal{E}_{0}/2}.  \label{BSeqnIFGbased}
\end{equation}%
This is precisely Eq. (7-7) of Ref. \cite{Schrieffer} (where all energies
are measured from the Fermi level) and in slightly different form than Eq.
(33.2) of Ref. \cite{AGD} \textit{before }assuming that $\mathcal{E}_{0}$ is
pure imaginary. If we assume that $\mathcal{E}_{0}$ can be real and $<0$
this eigenvalue refers to the 2e-CP \textquotedblleft
sector\textquotedblright\ in the BCS model interaction. In this case, the
denominator in the first term never vanishes since the pole lies below $%
E_{F} $. A similar argument holds on considering the 2h-CP sector when $%
\mathcal{E}_{0}>0$ and now ignoring the first integral.

However, if we consider both particles and holes simultaneously as in (\ref%
{BSeqnIFGbased}), then we must take into account that,\ if there exists a
real binding energy, there is a pole in one or the other integration
intervals depending on the sign of $\mathcal{E}_{0}$. If one assumes that $%
\mathcal{E}_{0}/2=i\alpha ,$ as in \cite{Schrieffer} or in \cite{AGD} with $%
\alpha $ real both integrals are now free of singularities and can be
integrated directly to give 
\begin{equation}
1=\frac{\lambda }{2}\left\{ \ln \frac{\hbar \omega _{D}-i\alpha }{-i\alpha }%
-\ln \frac{-i\alpha }{-\hbar \omega _{D}-i\alpha }\right\} =\frac{\lambda }{2%
}\ln \left[ \frac{\hbar ^{2}\omega _{D}^{2}+\alpha ^{2}}{\alpha ^{2}}\right].
\end{equation}
This yields the well-known pair of purely-imaginary roots reported in Refs. 
\cite{Schrieffer,AGD,bts}, namely%
\begin{equation}
\mathcal{E}_{0}^{BS}=\pm i2\hbar \omega_{D}/\sqrt{\exp (2/\lambda )%
{- 1}} \quad \mathrel{\mathop{\longrightarrow}\limits_{%
\lambda \rightarrow 0}} \quad \pm i2\hbar \omega _{D}\exp (-1/\lambda
)\equiv \pm i\lim_{\lambda \rightarrow 0}\Delta  
\label{2e2hCPIFG}
\end{equation}%
where%
\begin{equation}
\Delta =\hbar \omega _{D}/\sinh (1/\lambda)
\label{BCSDelta}
\end{equation}%
is the zero-temperature BCS energy gap \cite{bcs}.

A more general solution $\mathcal{E}_{0}$\ can be obtained by assuming,
without loss of generality, that it be \textit{complex}, namely%
\begin{equation}
\mathcal{E}_{0}\equiv r\exp i\phi  \label{E_0}
\end{equation}%
in (\ref{BSeqnIFGbased}). Then, direct integration in both terms of (\ref%
{BSeqnIFGbased})\ yields%
\begin{equation}
\frac{2}{\lambda }=\ln \frac{\mathcal{E}_{0}^{2}-(2\hbar \omega _{D})^{2}}{%
\mathcal{E}_{0}^{2}}\equiv \ln [\rho \exp i\theta ]=\ln \rho +i\theta.
\label{sol}
\end{equation}
Defining 
\begin{equation}
(2\hbar \omega _{D})^{2}/r^{2}\equiv \beta ^{2}>0  
\label{alfa}
\end{equation}%
and equating real and imaginary parts of (\ref{sol}) leads to%
\begin{equation}
\mbox{a)} \quad \rho = \sqrt{1+\beta ^{4}-2\beta ^{2}\cos 2\phi} \qquad %
\mbox{and} \qquad \mbox{b)} \quad \theta = \tan ^{-1}\frac{\beta ^{2}\sin
2\phi }{1-\beta ^{2}\cos 2\phi}.  
\label{rhotheta}
\end{equation}%
Clearly, (\ref{sol}) means that%
\begin{equation}
2/\lambda =\ln \rho \quad \quad \mbox{and} \quad \quad 0 = \theta .
\label{eqlamtheta}
\end{equation}%
The last identity substituted in (\ref{rhotheta}b) implies that $\beta
^{2}\sin 2\phi =0$\ which in turn is satisfied for%
\begin{equation}
\phi =0, \, \pm \pi /2,\, \pm \pi ,\, \pm 3\pi /2,\cdots .  
\label{phivalues}
\end{equation}

Consider the solutions with $\phi =\pm \pi /2$ so that (\ref{rhotheta}a)
becomes $\rho =\sqrt{1+\beta ^{4}+2\beta ^{2}}=1+\beta ^{2}>0.$ Thus, the
first relation in (\ref{eqlamtheta}) exponentiated gives $\exp (2/\lambda
)=1+\beta ^{2}$\ and again recalling the definition (\ref{alfa}) for $\beta $
as well as (\ref{E_0}) leads to $\exp (-i\phi )\mathcal{E}_{0}\equiv
r=2\hbar \omega _{D}/\sqrt{\exp (2/\lambda )-1}$\ whereupon inserting $\phi
=\pm \pi /2$ gives precisely (\ref{2e2hCPIFG}).

However, if we take the first solution $\phi =0$ of (\ref{phivalues}) in (%
\ref{rhotheta}a) this becomes $\rho =\sqrt{(1-\beta ^{2})^{2}}=\left\vert
1-\beta ^{2}\right\vert .$\ Since now $\mathcal{E}_{0}\equiv r$ and
recalling (\ref{alfa}) one obtains the \textit{real} energy eigenvalues%
\begin{equation}
\mathcal{E}_{0}^{BS}\,\mathcal{=\pm }\frac{2\hbar \omega _{D}}{\sqrt{\exp
(2/\lambda )+1}}\quad \mathrel{\mathop{\longrightarrow}\limits_{\lambda
\rightarrow 0}}\quad \pm 2\hbar \omega _{D}\exp (-1/\lambda )\equiv \pm
\lim_{\lambda \rightarrow 0}\Delta .  
\label{e0real}
\end{equation}%
In magnitude, this is always much larger than (\ref{ordCP}) since $%
e^{-1/\lambda }\gg e^{-2/\lambda }$ as $\lambda \rightarrow 0$. In Fig. 2
we plot the ratio $\mathcal{E}_{0}^{BS}/\mathcal{E}_{0}^{C}$ of the exact (%
\ref{e0real}) to the exact (\ref{ordCP}) (upper curve); the ratio $\Delta /%
\mathcal{E}_{0}^{BS}$ of exact BCS gap $\Delta $\ (\ref{BCSDelta}) to real
BS binding energy $\mathcal{E}_{0}^{BS}$ found here (lower full curve);
dashed curve is ratio of exact $\Delta $ to its weak-coupling limit, rhs of (%
\ref{BCSDelta}). Even for $\lambda $ as large as $1/2$ (the Migdal upper
limit \cite{Migdal}, marked as thin vertical line; see also Ref. \cite{Blatt}
p. 204.) this ratio is only about $1.019$. Successive values $\pm \pi ,$ $%
\pm 3\pi /2,\cdots $\ for $\phi $ in (\ref{phivalues}) give nothing new.
Solutions (\ref{2e2hCPIFG}) are the well-known imaginary roots of the BS
integral equation in the ladder approximation; they have been reported in
Refs. \cite{Schrieffer, AGD, bts}. The purely \textit{real} energies (\ref%
{e0real}) found here appear to be new. 
\begin{figure}[tbh]
\centerline{\epsfig{file=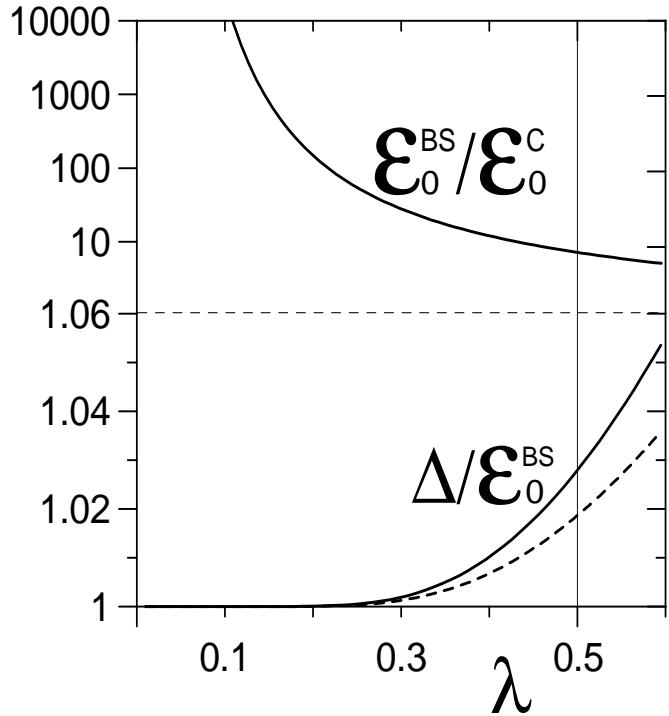,height=3.8in,width=3.6in}}
\caption{ Bottom: Ratio of exact BCS gap $\Delta $\ (\ref{BCSDelta})
to real BS binding energy found here (lower full curve). Dashed curve is
ratio of exact $\Delta $ to its weak-coupling limit, extreme rhs of (\ref%
{BCSDelta}). Upper full curve is ratio of \textit{real} exact $\mathcal{E}%
_{0}^{BS}$\ (\ref{e0real}) to exact $\mathcal{E}_{0}^{C}$ (\ref{ordCP}).
Horizontal thin dashed line marks drastic scale change. Vertical thin line
is the Migdal upper limit \cite{Migdal} on $\lambda$.}
\end{figure}

In summary, we find real solutions for the binding energy of a
\textquotedblleft generalized\textquotedblright\ Cooper pair when the
underlying BCS-type interaction is allowed to act between pairs of holes as
well as of electrons through a Bethe-Salpeter equation that restores
particle-hole symmetry around the Fermi level. The magnitude of the binding
energy coincides with the BCS energy gap in the weak coupling regime.
Finally, we note that the correct physical CP binding energies $\pm 2\Delta $
instead of $\pm \Delta $,\ follow when the ideal-Fermi-gas sea is replaced
by a BCS-correlated sea, as in Ref. \cite{Honolulu} Eqs. (12) and (13).

\noindent \textbf{Acknowledgments} { MF, MdeLl and MAS acknowledge
UNAM-DGAPA-PAPIIT (Mexico) for grants IN106401 \& IN114708, and CONACyT
(Mexico) grants 41302F and 43234F, for partial support.}\newline


\begin{thebibliography}{99}

\bibitem{BS} E.E. Salpeter and H.A. Bethe, Phys. Rev. \textbf{84}, 1232
(1951).

\bibitem{FW} A.L. Fetter and J.D. Walecka, \textit{Quantum Theory of
Many-Particle Systems} (McGraw-Hill, New York, 1971).

\bibitem{Honolulu} M. Fortes, M.A. Sol\'{\i}s, M. de Llano, and V.V.
Tolmachev, Physica C \textbf{364,} 95 (2001).

\bibitem{BG} H.A. Bethe and J. Goldstone, Proc. Roy. Soc. (London) \textbf{A
238}, 551 (1957).

\bibitem{Coo} L.N. Cooper, Phys. Rev. \textbf{104}, 1189 (1956).

\bibitem{BCSandCPs} M. de Llano, F.J. Sevilla, and S. Tapia, Int. J. Mod.
Phys. B \textbf{20}, 2931 (2006).

\bibitem{PRBcomm} M. de Llano and J.J. Valencia, Mod. Phys. Lett. B \textbf{%
20}, 1067 (2006).

\bibitem{Schrieffer} J.R. Schrieffer, \textit{Theory of Superconductivity}
(Benjamin, Reading, MA, 1983) p. 168.

\bibitem{AGD} A.A. Abrikosov, L.P. Gorkov, and I.E. Dzyaloshinskii, \textit{%
Methods of Quantum Field in Statistical Physics }(Dover, NY, 1975) \S\ 33.

\bibitem{bts} N.N. Bogoliubov, V.V. Tolmachev, and D.V. Shirkov, \textit{A
New Method in the Theory of Superconductivity} (Consultants Bureau, NY,
1959) p. 44.

\bibitem{bcs} J. Bardeen, L.N. Cooper, and J.R. Schrieffer, Phys. Rev. 
\textbf{106}, 162 and \textbf{108}, 1175 (1957).

\bibitem{Migdal} A.B. Migdal, JETP \textbf{7}, 996 (1958).

\bibitem{Blatt} J.M. Blatt, \textit{Theory of Superconductivity} (Academic,
New York, 1964).
\end{thebibliography}
\end{document}